\documentclass{aa}
\usepackage{graphicx}
\newcommand{\be}{\begin{equation}}
\newcommand{\ee}{\end{equation}}
\begin{document}

\title{The radio surroundings of the microquasar GRO J1655$-$40}

\author{J.A. Combi$^{1}$, G.E. Romero$^{1}$, P. Benaglia$^{1}$,
I.F. Mirabel$^{2, 3}$}

\offprints{J.A. Combi}

\institute{$^1$ Instituto Argentino de Radioastronom\'{\i}a,
C.C.5, (1894) Villa Elisa, Buenos Aires, Argentina \\ $^2$
CEA/DMS/DAPNIA/Service D'Astrophysique, Centre d'Etudes de Saclay,
F-91191 Gif-sur-Yvette, France \\$^3$ Instituto de Astronom\'{\i}a
y F\'{\i}sica del Espacio (IAFE), C.C. 67, Suc. 28, Buenos Aires,
Argentina}

\date{Received 12 February 2001 / Accepted 22 February 2001 }
\abstract{ We report the results of a study of the radio
surroundings of the superluminal microquasar GRO J1655$-$40. We
have searched for extended continuum structures that might be
indicative of the presence of a supernova remnant (SNR) associated
with the formation of the compact object in the binary system. We
also carried out HI-line observations of the region looking for a
local minimum created by an explosive event. Our results indicate
that there is, in fact, a bubble in the large-scale HI
distribution around GRO J1655$-$40. We suggest that this structure
might be created by the original supernova explosion occurred a
few hundred thousand years ago and whose signatures can be traced
by the overabundance of $\alpha$-elements recently found by
Israelian et al. (1999) in the companion star. }

\titlerunning{Radio surroundings of GRO J1655$-$40}
\authorrunning{J.A. Combi et al.}

\maketitle


\section{Introduction}
The soft X-ray transient GRO J1655$-$40 is one of the best black
hole candidates in the Galaxy (Mirabel \& Rodriguez 1999). It is a
low mass X-ray binary whose compact object has a mass in the range
$5.5-7.9$ $M_{\odot}$ (Shahbaz et al. 1999). The system is located
at 3.2 kpc from the Earth and presents frequent superluminal
ejections at radio wavelengths (e.g. Tingay et al. 1995, Hjellming
\& Rupen 1995).

Recently, Israelian et al. (1999) found convincing evidence of a
supernova origin for the black hole in GRO J1655$-$40. They
detected a clear overabundance of $\alpha$-elements in the
secondary star, whose internal temperatures are not high enough as
to synthesize them. These elements seem to have been produced in a
massive companion (with $25-40$ $M_{\odot}$) which exploded both
contaminating the secondary star with matter enriched with
nucleosynthetic products and creating the black hole. Israelian et
al. (1999) estimated that the explosion should have occurred less
than $10^6$ years ago.

In this {\sl Letter}, we present the results of a study of the
radio environment of GRO J1655$-$40. Our aim is to detect any
existent signature of an explosive event in this region. A
supernova explosion should have introduced modifications in the
ISM around the binary system, displacing material and creating a
low density bubble. Additionally, the expanding shock front should
accelerate electrons up to relativistic energies, producing a
shell-type source in the radio continuum. If the age of the system
is less than $10^6$ years, these observational features might be
detectable.

\section{Data analysis and new observations}

\subsection{Radio continuum}

We have used data from the 4.85-GHz PMN survey (Condon et al.
1993) in order to produce a large-scale map ($\sim 1^{\circ}
\times 1^{\circ}$) of the surroundings of GRO J1655$-$40. Although
these data are optimized for sizescales of less than 40
arcminutes, the good resolution and sensitivity of the images
render them a valuable tool to search for SNR candidates,
especially if they are not of large angular size (Duncan et al.
1997). For this work, we apply to the 4.85-GHz data an additional
filtering process in order to remove the galactic diffuse emission
on scales larger than 30 arcminutes (see Combi et al. 1998 for
details of the Gaussian filtering method). The resulting map is
shown in Figure 1, along with a 1.4-GHz image of the small-scale
emission of the inner region obtained from the NVSS Sky Survey
(Condon et al. 1998) at a better resolution. These latter
observations were obtained with the Very Large Array (VLA) in the
compact D and DnC configurations.

\begin{figure}
\caption{{\rm Upper panel:} Filtered radio emission at 4.85 GHz of
the region surrounding GRO J1655$-$40. The position of the
microquasar is marked by a cross. Contours are labeled in steps of
10 mJy beam$^{-1}$, starting from 35 mJy beam$^{-1}$. {\rm Lower
panel:} Small-scale, higher resolution, VLA image at 1.4 GHz for
the region close to GRO J1655$-$40. Radio contours are in steps of
2 mJy beam$^{-1}$ starting from 1 mJy beam$^{-1}$. The arrows
indicate the directions of the observed superluminal ejections.}
\label{fig.1}
\end{figure}

\subsection{HI observations}

We have performed HI observations towards the area of interest
with a 30-m single dish telescope located at the Instituto
Argentino de Radioastronom\'{\i}a (IAR), Villa Elisa, Argentina.
The observations were carried out during four consecutive sessions
on November 20-23, 2000. The receiver is a helium-cooled HEMT
amplifier with a 1008-channel autocorrelator at the backend.
System parameters and additional details of the observational
technique can be found in Combi et al. (1998). The HI line was
observed in hybrid total power mode and the sky was sampled on a
$0.35^{\circ}$ rectangular grid. Each grid position was observed
during 60~s with a velocity resolution of \mbox{$\sim 1$ km
s$^{-1}$} and a coverage of $\pm 450$ km s$^{-1}$. A set of HI
brightness temperature maps ($\Delta T_{\mbox{\scriptsize rms}}
\sim 0.2$ K) were made for the velocity interval $\Delta v=$($-80$
km s$^{-1}$, $+20$ km s$^{-1}$). Those maps for the interval $-42$
to $-30$ km s$^{-1}$, corresponding to the HI distribution at a
distance centered at $\sim 3.5$ kpc, are shown in Figure 2.

\section{Main results}

Figure 1 (upper panel) shows the radio continuum image at 4.85 GHz
of the surroundings of GRO J1655$-$40. A small ($\sim 24'\times
24'$) semi-circular radio source, which very much resembles a SNR,
can be clearly seen in the center of the image. This extended,
partial shell source is centered at
$(l,b)\approx(345.14^{\circ},+2.48^{\circ})$. It has an integrated
flux density of $\sim 0.5$ Jy at 4.85 GHz, although uncertainties
due to the background filtering make difficult to estimate a
realistic value. The microquasar is located at a distance of
$\sim0.15$ deg from the center of the radio source.

Figure 1 (lower panel) shows the VLA image at 1.4 GHz (Condon et
al. 1998) of the vicinity of GRO J1655$-$40. The microquasar was
undetectable at the epoch of the observations and its position is
marked with a cross. There are seven point sources near the
microquasar with fluxes above 5 mJy. They have been labeled from
S1 to S7 in the image (notice that the SNR candidate is not
visible in this map because these data are optimized for point
source detection). The bright, compact source located at $(l,b)
\sim 345.1^{\circ}$, $+2.37^{\circ}$ is identified in the PMN
catalogue as PMN J1654-3949. The sources S1 to S6 have no entry in
any point source catalog at present. The measured characteristics
of all these sources are summarized in Table 1.

Figure 2 clearly shows the development of a cavity around GRO
J1655$-$40 in the HI maps from $-42$ to $-30$ km s$^{-1}$.
Although the cavity do not form a shell due to the steep density
gradient towards the galactic plane, the existence of a local
minimum in the HI distribution at the position of the microquasar
is evident. This kind of ``horse-shoe" morphology has been
observed around early-type stars with very strong stellar winds
(e.g. Benaglia \& Cappa 1999), where a strong energy release in a
highly inhomogeneous medium has occurred.

In Figure 3 we present the integrated column density map for the
velocity interval from $-40$ to $-30$ km s$^{-1}$. Standard
galactic rotation models (Fich et al. 1989) indicate that this HI
hole is located at a kinematic distance of 3 to 4 kpc, in good
accordance with the estimated distance to GRO J1655$-$40.

\begin{figure}
\caption{HI brightness temperature channel maps (contour labels in
K) obtained for the velocity range from $-42$ to $-30$ km s$^{-1}$
around the position of GRO J1655$-$40 (indicated with a cross).
Due to the differential galactic rotation the expected velocity of
the interstellar medium at the distance of 3.2 kpc of the X-ray
binary is $\sim$ 38 km s$^{-1}$} \label{fig.2}
\end{figure}
\begin{table}
\caption[]{Point radio sources near GRO J1655$-$40}
\begin{flushleft}
\begin{tabular}{l c c c}
\noalign{\smallskip} \hline Source & $(l,b)$     &  $F_{\rm 1.4
\;GHz}$ & ID  \cr
       & (deg, deg)  &  (mJy)         &     \cr
\hline S1     & ($344.87^{\circ}$, $+2.62^{\circ}$)& 24.9 & -- \cr
S2     & ($344.92^{\circ}$, $+2.54^{\circ}$)& 12.8 & -- \cr S3 &
($344.96^{\circ}$, $+2.46^{\circ}$)& 5.43 & -- \cr S4     &
($344.98^{\circ}$, $+2.42^{\circ}$)& 4.48 & -- \cr S5     &
($345.02^{\circ}$, $+2.38^{\circ}$)& 37.7  & --\cr S6     &
($345.03^{\circ}$, $+2.49^{\circ}$)& 5.83  & -- \cr S7     &
($345.10^{\circ}$, $+2.37^{\circ}$)& 10.6  & PMN J1654-3949 \cr
\hline
\end{tabular}
\end{flushleft}
\end{table}

\begin{figure}
\caption{Integrated column density map for the velocity range from
$-40$ to $-30$ km s$^{-1}$. Contour labels are in units of
$10^{19}$ cm$^{-2}$. Microquasar position indicated with a cross.}
\label{fig.3}
\end{figure}

\section{Discussion}

If we assume that the extended radio structure shown in Fig. 1,
upper panel, is a supernova remnant located at the same distance
than the microquasar, we get that its radius should be $\sim 15$
pc. The surface brightness of the source results to be abnormally
low for such a remnant ($\Sigma_{\rm 4.8\; GHz}\sim
1.3\times10^{-22}$ W m$^{-2}$ Hz$^{-1}$ sr$^{-1}$). On the other
hand, the microquasar is displaced about $0.15^{\circ}$ from the
center of the semi-shell source. Although the transverse velocity
of GRO J1655$-$40 is not known, it is constrained to be less than
300 km s$^{-1}$ (Brandt et al. 1995). At this velocity, it would
have taken more than $2.7\times10^4$ yr to reach its current
position. And for a velocity similar to the observed radial
velocity ($\sim 114$ km s$^{-1}$), the age of the system would be
$\sim 82\,500$ yr. These timescales are incompatible with the size
of the remnant and our knowledge of the ambient density in this
region (see below), and lead us, consequently, to discard the
extended radio source in Fig. 1 as possibly associated with GRO
J1655$-$40. This source could be a background SNR or, perhaps, an
artifact formed by several close and unresolved point sources.

At larger sizescales there are not other clear continuum
structures that could be associated with a SNR. We have inspected
larger maps, of $\sim 6^{\circ}\times6^{\circ}$, in search of
additional signatures, but nothing was found. This is a very
confused region located nearly towards the Galactic Center and the
detection of weak continuum features with low surface brightness
is extremely difficult. Even at small scales there is a
significant number of sources, as it is shown in Fig. 1, lower
panel. Here, the two closest sources to the microquasar, S3 and
S6, seem to be at approximate equal angular distances, in opposite
directions, of the X-ray binary. However, it is unlikely that they
could be old emission knots ejected by the central object, because
they are on an axis displaced $30^{\circ}$ from the axis
determined by the known ejections (Hellming \& Rupen 1995), which
is also shown in the figure.

Contrary to the often misleading continuum data, the HI
observations can provide more clear signatures of an explosive
event in this region, because the differential rotation motion of
the material allows a glimpse of the matter distribution along the
line of sight. The channel maps in Fig. 2 show the development in
the velocity distribution of a cavity in the HI with a
``horse-shoe" shape and an average radius of $\sim75$ pc at a
distance of 3.2 kpc. The integrated column densities shown in Fig.
3 can be used to estimate the mass displaced to form the cavity.
Assuming that the background density is given by the density at
the minimum, we estimate that at least 25 000 $M_{\odot}$ have
been removed from this region. Since the average expansion
velocity of the shell is $\sim 15$ km s$^{-1}$, we obtain that a
lower limit to the original energy release which started the mass
motion at the center of the cavity is $\sim 6\times10^{49}$ erg.
The expansion velocity should have been considerably higher in the
past, in such a way that the real value is probably in the range
$10^{50}-10^{51}$ erg, what is quite consistent with the supernova
hypothesis. This is reinforced by the fact that there are not
known early-type stars with stellar winds strong enough to create
such a bubble in this sector of the sky. It is noteworthy that
even higher values for the initial energy release are possible if
the explosion was caused by a hypernovae event as recently
proposed by Brown et al. 2000.

The mean number density in this region can be obtained averaging
the column densities within the bubble. This provides a rough
estimate of the average density of the ISM before the explosion.
We obtain a value $n_0\sim 0.4$ cm$^{-3}$. This density and the
size of the cavity ($R\sim75$ pc) indicate that the SNR should be
well within the radiative phase. An exact estimate of the time
elapsed since the explosion is not possible without a knowledge of
the detailed structure of the ISM in this region, but simulations
with an original energy release of few times $10^{50}$ erg and
some reasonable conditions for the cloud properties suggest ages
in the range $10^5-10^{5.5}$ yr (Cowie et al. 1981). The continuum
emission of the remnant, with sizescales of several degrees,
should have been removed during the filtering of the background
radiation in the PMN survey. New observations with good
sensitivity and single dish telescopes over a large area are
necessary to detect the SNR radio structure and help to make a
more accurate estimate of its age, if the remnant is not already
dissipated.

In summary, we have presented radio continuum and HI maps of the
surroundings of the black hole candidate GRO J1655$-$40. A bubble
centered at the microquasar has been detected in our HI data. The
energy necessary to create this bubble and its approximate age are
consistent with the hypothesis that it is the trace left in the
ISM by the supernova explosion occurred when the black hole was
formed.

\begin{acknowledgements}
This work was supported by CONICET (PIP 0430/98), ANPCT (PICT
03-04881) and Fundaci\'on Antorchas.
\end{acknowledgements}

{}


\begin{thebibliography}{}

\bibitem{} Benaglia P., Cappa C.E. 1999, A\&A 346, 979
\bibitem{} Brandt W.N., Podsiadlowski Ph., Sigurdsson S. 1995,
MNRAS 277, L35
\bibitem{} Brown G.E., Lee C.-H., Wijers R.A.M.J., Lee H.K., Israelian G.,
Bethe H.A. 2000, New Astron. 5, 191.
\bibitem{} Combi J.A., Romero G.E., Benaglia P. 1998, A\&A 333, L91
\bibitem{} Condon J.J., Griffith M.R., Wright A.E. 1993, AJ 106,
1095
\bibitem{} Condon J.J., Cotton W.D., Greisen E.W., et al. 1998, AJ 115,
1693
\bibitem{} Cowie L.L., McKee C.F., Ostriker J.P. 1981, ApJ 247,
908
\bibitem{} Duncan A.R., Stewart R.T., Haynes R.F., Jones K.L.
1997, MNRAS 287, 722
\bibitem{} Fich M., Blitz L., Stark A.A. 1989, ApJ, 342, 272
\bibitem{} Hjellming R.M., Rupen M.P. 1995, Nat 375, 464
\bibitem{} Israelian G., Rebolo R., Basti G., Casares J., Martin E.L.
1999, Nat 401, 142
\bibitem{} Mirabel I.F., Rodr\'{\i}guez L.F. 1999, ARA\&A 37, 409
\bibitem{} Shahbaz T., van der Hooft F., Casares J., Charles P.A.,
van Paradijs J. 1999, MNRAS 306, 89
\bibitem{} Tingay S.J., et al. 1995, Nat 374, 141

\end{thebibliography}
\end{document}